# Spherulite-enhanced Macroscopic Polarization in Molecular Ferroelectric Films from Vacuum Deposition

Bibek Tiwari[1], Yuanyuan Ni[1], Jackson Savage[2], Ellen Daugherty[3], Bharat Giri[1], Xin Li[1], and Xiaoshan Xu[1,4*]

[1]*Department of Physics and Astronomy, University of Nebraska-Lincoln, Nebraska 68588, USA*

[2]*Department of Physics, University of Chicago, Illinois 60637, USA*

[3]*Department of Chemistry, The College of Wooster, Ohio 44691, USA*

[4]*Department of Physics and Astronomy and the Nebraska Center for Materials and Nanoscience, University of Nebraska-Lincoln, Lincoln, Nebraska 68588, USA*

*xiaoshan.xu@unl.edu

**Abstract:**

Proton-transfer type molecular ferroelectrics hold great application potential due to their large spontaneous polarizations, high Curie temperatures, and small switching fields. However, it is puzzling that preparation of quasi-2D films with macroscopic ferroelectric behaviors has only been reported in few molecular ferroelectrics. To resolve this puzzle, we studied the effect of microstructures on macroscopic ferroelectric properties of 5,6-Dichloro-2-methylbenzimidazole (DC-MBI) films grown using low-temperature deposition followed by restrained crystallization (LDRC) method. We revealed a competition between dense spherulites and porous microstructures containing randomly oriented nanograins in as-grown films. Post-growth annealing at moderate temperature promotes the formation of spherulites which leads to macroscopic ferroelectric polarization switching. These results highlight microstructure density as a critical factor for macroscopic ferroelectric properties, potentially resolving the puzzle for absence of macroscopic ferroelectric behavior in molecules ferroelectric films. We expect the approach for enhancing microstructure density offered in this work to greatly advance fabrication of quasi-2D molecular ferroelectrics films and to unlock their potential in device applications.

**Keywords:** Organic Ferroelectrics, Hydrogen-bonded, DC-MBI, Physical vapor deposition, thin film

# I. Introduction

The potential advantages of utilizing organic ferroelectric materials have significant implications for a wide range of applications. Specifically, their outstanding dielectric, pyroelectric, and piezoelectric properties, comparable to those found in traditional inorganic ferroelectrics makes them promising candidates for applications such as actuators, transducers and pyroelectric detectors[1–4]. Their environmentally friendly characteristics contrast with the prevalent use of lead and rare-metal ferroelectrics in today's applications.

Proton-transfer type ferroelectrics of molecular crystals, compared with other organic ferroelectrics, have the advantage of low coercivity (~10 kV/cm) and large spontaneous polarizations[5]. In particular, croconic acid exhibits spontaneous polarizations around 30 $\mu$C/cm$^2$,[6] which is on par with prototype inorganic ferroelectric BaTiO$_3$. Because hydrogen bonds are responsible for cohesion in proton-transfer molecular ferroelectrics, spontaneous polarization often persists up to the melting temperature, which is typically above 400 K[5,7,8], also comparable to the Curie temperature of BaTiO$_3$.

The application potential of organic ferroelectrics hinges on viable thin film fabrication approaches that retain the ferroelectric properties. On the other hand, growth of quasi-2D films of molecular ferroelectrics appears to conflict with the crystallization process[9,10], which is required for ferroelectric order, since the latter tend to form 3D structures given the weak interactions between the molecular crystals and most inorganic substrates. The recently established method of low-temperature deposition followed by restrained crystallization (LDRC) has offered a route to control film crystallization while maintaining the quasi-2D morphology. The LDRC method was applied to methyl-benzimidazole (MBI)[11], resulting in quasi-2D highly-oriented dense spherulites which demonstrate single-crystal-level ferroelectric properties. On the other hand, croconic acid films often consist of porous microstructures of randomly oriented nanograins[12]. Although ferroelectric polarization switching was routinely shown for individual grains[13,14], it was rarely demonstrated with macroscopic electrodes for croconic acid films[15]. Therefore, in addition to the quasi-2D morphology and crystallization, what's the optimal microstructure for ferroelectric behavior and how to achieve the optimal microstructure, are critical questions.

To answer these questions, we focus on 5,6-Dichloro-2-methylbenzimidazole (DC-MBI), a proton-transfer type ferroelectric with spontaneous polarization of about 10 $\mu$C/cm$^2$ (at 373 K)[6,16]. As shown in **Fig. 1(a)**, DC-MBI crystallizes in a Pca2$_1$ structure in which the molecules form a chain-like structure connected by the N-H··N hydrogen bonds[6]. The spontaneous polarization is along the *c* axis which is the chain direction. The melting point of DC-MBI is about 250 °C, which is in between that of MBI ($\approx$ 175 °C) and that of croconic acid ($\approx$ 300 °C)[17–19]. Therefore, the DC-MBI films are expected to have microstructure characteristics of both MBI (dense spherulites) and croconic acid (porous randomly oriented nanograins). Indeed, using the LDRC method [**Fig. 1(b)**], we observe a competition between spherulites and porous microstructure of randomly oriented nanograins. Post-deposition annealing at moderate temperature promotes the formation of spherulites which leads to well-defined ferroelectric switching loops and enhanced dielectric constants.

## II. Results and Discussions

**Growth of DC-MBI films**

We first determine the effect of substrate temperature on the formation of spherulites. **Figs. 2(a-f)** show the laser microscopy images for the DC-MBI films (for powder sample morphology via SEM and TEM, see **Fig. S1**) deposited on sapphire substrates in high vacuum at various substrate temperatures $T_s$ with nominal thickness 1 μm, followed by slowly warming up (~0.2 K per minute) to room temperature. For $T_s$ = 294 K [**Fig. 2(a)**], the most salient features in the image are the disk-shaped features (indicated by the box), with diameter ≈ 100 μm and thickness 1.0 ± 0.6 μm (for features statistics at other $T_s$ see Table 1 on supplementary). We call them Type-I features. In addition, dendric-shaped features are also visible. The large white area of this image, on the other hand, is not covered by the film. For $T_s$ ≤273 K, besides the Type-I feature, another disk-shaped feature with thickness 2.0 ± 1.2 μm emerges, marked as Type-II in **Fig. 2(b)**. Below $T_s$ = 173 K, the Type-I and Type-II features can still be identified, but the substrates are fully covered, as indicated by the large grey area. **Fig. 2(g)** shows the area coverage as functions of thickness. Overall, coverage of the Type-I and Type-II features both diminishes for low and high $T_s$, although the maximum coverage for the Type-II feature occurs for higher $T_s$ than that of the Type-I feature.

The Type-I feature appears to be spherulites. As shown in **Fig. 2(h)**, which is the closed-up view of the Type-I feature in **Fig. 2(b)**, on top of the round shape, there are also ring-like bands around the center evenly spaced along the radial direction, which is a typical trait of spherulites from the rhythmic growth[20,21]. The Type -II feature in **Fig. 2(i)**, although also circular, appears to be more porous with fluffy boundaries without the rings, which is consistent with spherical dendrites[22]. The more ordered nature of the Type-I feature is confirmed by the scanning electron microscopy (SEM) images shown in **Fig. S2**. In contrast, the background grey area in **Fig. 2(f)** shows no sign of order.

The $T_s$ dependence of the images in **Fig. 2(a-f)** can be understood in terms of spherulite/dendrite nucleation. Below $T_s$ = 173 K, the sticking coefficient[23] is high for the deposition, which leads to the full coverage of the films. Nucleation of spherulites and dendrites occurs when the films are warmed up to room temperature. The disk shape is the result of the constraint imposed by the film morphology due to the low temperature deposition[11]. Above $T_s$ = 173 K, the sticking coefficient is reduced, which means nucleation of spherulites, and dendrites is important for the film coverage. Normally, growth of spherulite requires large supersaturation and slow growth speed[21] compared with that of the dendrites, which is consistent with the lower $T_s$ for maximum coverage of the spherulite compared with that of the spherical dendrites in **Fig. 2(g)**. To promote the growth of spherulites, we decided that the optimal substrate temperature is $T_s$ =183 K.

Next, we study the effect of film thickness for $T_s$ =183 K. In order to measure the ferroelectric properties, the substrates were pre-patterned with Pt interdigital electrodes (IDE). **Figs. 3(a-d)** display the optical images for films on IDE (similar morphology observed on Au/Si substrate, **Fig. S3**) of various thickness with crossed polarizers, where the light path is the 1st polarizer, the film, and the 2nd polarizer. When the film is optically isotropic, no light goes through the 2nd polarizer, resulting in dark color in the image. On the other hand, if the film has in-plane anisotropy, it may rotate the light polarization which leads to non-zero transmission through the 2nd polarizer, unless the polarization of the light coincides with the optical axis of the film. Spherulites consists of fiber crystallites aligned along the radial direction. The optical axes are mostly along and perpendicular to the fiber respectively, meaning that light rotation is expected except when the radial direction aligns with either polarizer's axis. This is the origin of the Maltese cross, a hallmark trait of spherulites[21] as clearly observed for the Type-I feature in **Figs. 3(a-d)**. Here the axes of the maximum intensity are along the direction 45° from both polarizers' axes. For the Type-II feature, substantially weaker light intensity than that of the Type-I feature, without the Maltese cross, is observed, which is consistent with the less-ordered structure. The background area, called Type-III feature, shows minimal light intensity, consistent with a very disordered nature. Further the XPS measurement (**Fig. S4)** qualitatively shows the similar bonding environment for the powder and films.

The microstructure order of these features was confirmed by atomic force microscopy (AFM), as shown in **Figs. 3(e-g).** The step (≈ 200 nm) and the period (~10 um) of the IDE can be seen on the images. From Type-I to Type-III, the disorder increases. Type-I feature [**Fig. 3(e)**] consists of well-aligned crystallite fibers tightly packed together. In comparison, in the Type-II feature [**Fig. 3(f)**], the fibers are less aligned, forming a porous structure, but there is still a discernable alignment direction in the image. For the Type-III area [**Fig. 3(g)**], the fibers are randomly oriented, causing an even larger porosity; this makes the area optically isotropic, consistent with the minimal light intensity for the Type-III feature observed in **Figs. 3(a-d)**. The disorder and porosity can also be inferred from the roughness of the AFM images measured within the digits of the IDE, where the root-mean-square roughness is about 30 nm, 70 nm, and 100 nm for the Type-I, Type-II, and Type-III regions respectively. Therefore, a clear correlation can be found between the brightness in the crossed-polarizer optical images, the microstructure density/porosity, and the order of the crystallite fibers.

The thickness dependence for the coverage of the different types of features is shown in **Fig. 3(h)**. As the thickness increases, the coverage of the Type-II feature increases monotonically, which is accompanied by the monotonic reduction of the Type-III area. In particular, at 4 μm, most of the area is occupied by the Type-I and Type-II features, although the less ordered type-II feature dominates.

**Annealing and Electrical Study of the films**

According to the previous work on MBI[11], single-crystal-level ferroelectric properties were obseved in films consisting of spherulites. To increase the population of spherulites, which is the most ordered microstructures among the three types of feature, we carried out annealing on the film samples by slowly (~0.2 K per minute) warming up the samples from room temperature to 353 K and stay for 2.5 hours in the atmosphere. **Figs. 4(a&b)** are optical images (not the same area) with crossed polarizers of a 4 μm film before and after the annealing. After the annealing, most area in the image is covered by the Maltese cross, suggesting that the Type-I feature (spherulites) dominates. In other words, the coverage of the Type-II and Type-III features is largely reduced. At the same time, the sizes of the Maltese crosses after the annealing [**Fig. 4(b)**] is often much larger than those before the annealing [**Fig. 4(a)**] and closer to the sizes of the Type-II features in **Fig. 4(a)**, suggesting that the Type-II feature may obtained better fiber order and turned into spherulite like the Type-I feature. AFM images reveal more details on the effect of annealing. **Fig. 4(c)** shows an area of packed fibers in the Type-I feature. After the annealing [**Fig. 4(d)**], the fibers become apparently longer, probably due to merging. More importantly, the alignment (or order) becomes more obvious.

Specular θ-2θ x-ray diffraction (XRD) reveals the preferred crystalline orientations of the films (for powder and simulated comparison see **Fig. S1d**), which changes clearly with thickness, as shown in **Fig. 4(e)**. For the 1 μm film, with the weak diffraction intensity, the most well-defined peaks are (210) and (311). In addition, the (010) peak, which is almost invisible for the power sample, is also discernable. Therefore, for the 1 μm film, the (010) direction appears to be the preferred out-of-plane direction. The (010) peak becomes well-defined for the 2 μm and 3 μm films, although the relative intensity decreases with the film thickness; for the 4 μm film, the (010) peak disappears. In contrast, the (002) peak, which is also very weak in the powder sample, increases with the film thickness, suggesting that the (001) orientation is preferred for the thicker films. For the 4 μm film after annealing, the (002) peak still is strong, maintaining the (001) as the preferred orientation.

Therefore, annealing clearly improves the order of the crystallite fibers and density of the microstructure, which is expected to also improve the macroscopic ferroelectric properties of the DC-MBI films, as discussed below.

Polarization-voltage (P-V) relation measured in-situ reveals the effect of microstructure on the macroscopic ferroelectric properties. At $T_s$ = 183 K, the P-V relation is linear (for comparion at different thickness, see **Fig. S5**), indicating a linear capacitor behavior . In **Fig. 5(a)**, the capacitances are plotted as functions of thickness for various films; the analysis based on partial capacitance model[24] results in dielectric constants $\epsilon_{DC-MBI}$ = 2.7 for DC-MBI and $\epsilon_{Substrate}$ = 5.4 for the IDE. The value for DC-MBI is close to the value for amorphous MBI (2.93)[11], which is a factor of 30 less than the value for MBI single crystals. Therefore, DC-MBI films at $T_s$ =183 K is most likely in amorphous phase.

The effect of annealing (0.2 K per minute from room temperature to 353 K) can be viewed from **Fig. 5(b)** for a 4 μm film. Before annealing, the P-V relation at room temperature (294 K) has a visible but minimal opening, due to the limited order shown in **Fig. 5(b)**. Annealing at 353 K for 2.5 hours leads to a further opening of the loop, suggesting that the conversion to spherulites from more disordered microstructures leads to macroscopic ferroelectric behavior in DC-MBI (for local microscopic ferroelectric domain and poling results, see **Fig. S6 and Fig. S7**). After cooling down to room temperature (294 K), the P-V loop is still open, confirming that the P-V loop is mostly due to the irreversible process of microstructure ordering. The remanent polarization at 294 K after cooling is about 0.2 μC/cm², as measured using the PUND method at 0.5 Hz for the 4 μm thick sample.

The current-voltage (I-V) relation **Fig. 5(c)** along with the resistance measured at 150 V (**Fig. 6**) reveals a large change of resistance at transition temperature. During the deposition at $T_s$ = 183 K, the I-V relation is linear but with a small slope, corresponding to large (≈ 100 GΩ) resistance through the IDE. When the film is warmed up, the I-V relation ehxibits current peaks on top of a linear background. While the current peaks are consistent with the polarization switching, the slope of the linear background is the inverse of resistance, which decreases during the annealing procecess. At 353 K after 2.5 hours annealing, the resistance drops to about (≈ 1 GΩ), which is two orders of magnitude decrease compared with the value at 183 K for the amorphous phase. However, the subtle structrual transition happens at ~293K based from the capacitance, resistance and remenance measurement (**Fig. 6**) wherby the resistance drops to (≈ 7 GΩ). This large change of resistance is consistent with the resistance change during the amorphous to spherulite transition of MBI films. After cooling down to 294 K and below, the resistance remains at a small value (≈ 10 GΩ), again consistent with the irreversible microstructure ordering process.

**Discussion:**

The formation of spherulites in thin films requires two most important conditions: (1) large driving force, such as supercooling or super saturation, and (2) slow kinetics due to interface-controlled growth [21].The large driving force can be realized at low deposition temperature, which is also necessary for (1) reducing the diffusion so that a quasi-2D morphology can be formed and (2) increasing the sticking coefficient so that high film coverage can be achieved.

On the other hand, low deposition temperature may slow down the growth of spherulites to an unrealistic time scale. Therefore, post-deposition annealing is necessary. The optimal annealing temperature is determined by the crystal growth speed which reaches a maximum at an intermediate temperature $T_{max}$, because it is slow both at low temperature due to the slow diffusion and at high temperature due to the small driving force. Previous work indicates that $T < T_{max}$ is more favorable for the spherulite growth, while $T > T_{max}$ favors larger 3D single crystals[21]. Hence, the annealing temperature should satisfy $T < T_{max}$. In this work, we tried annealing at higher temperature (**Fig. S8**), but the results indicate growth of larger crystals or even re-evaporation of film materials which destroys the quasi-2D morphology.

The comprehensive microscopy technique used in this study, including optical microscopy, crossed-polarizer optical microscopy, SEM, and AFM, provides a clue of the how the macroscopic polarization of the films depends on the microstructure. At deposition temperature, the amorphous phase dominates, which is expected to exhibit a paraelectric behavior. After deposition and warming up to the room temperature, the microstructure of the crystalized phase is dominated by the porous spherical dendrites. The effect of porosity on macroscopic polarization can be demonstrated using a double layer model (**Fig. S9**)[25], which shows a greatly reduced dielectric constant and polarization compared with the single crystal. The post-deposition annealing is then the key to converting the more disordered porous microstructures to dense spherulites which are expected to exhibit dielectric constant and remanent polarization close to the single crystal. Here, the maximum $P_r$ found is about 0.2 μC/cm², which is still substantially lower than that of the single crystal value; this could be because the polar (001) axis is preferred along the out of plane direction [**Fig. 4(e)**], while the P-V measurements are in-plane.

Finally, not all materials are expected to grow as spherulite[21]. On the other hand, we can summarize more general rules for optimizing macroscopic ferroelectric properties (1) low-temperature deposition for quasi-2D morphology and amorphous phase, and (2) annealing at moderate temperature for controlled crystallization to reach dense microstructures without breaking the quasi-2D morphology. The importance of high microstructure density is the most critical finding of this work.

**Conclusion**

We have studied the growth of DC-MBI films using the LDRC method with various deposition temperatures, thickness, and post-deposition annealing. A mixed population of microstructures have been observed including dense spherulites, porous spherical dendrites, and even more porous randomly oriented fibers. With the deposition temperature optimized for spherulite populations, thicker films are dominated by spherical dendrites. Annealing at a moderate temperature converts spherical dendrites to spherulites, which exhibit macroscopic ferroelectric polarization switching. The relatively small remanent polarization is attributed to the preferred orientation of the polar axis along the out-of-plane direction while the measurements are in-plane. These results elucidate the importance of microstructure density in addition to the quasi-2D morphology in the fabrication of films for macroscopic ferroelectric properties. The approach established here, i.e., low-temperature deposition and controlled crystallization at moderate temperature is expected to be viable for most molecular ferroelectric films.

**Methods:**

We deposited micro-meter thick DC-MBI films using Angstrom Engineering's Physical Vapor Deposition (EvoVac) within a vacuum setting (~$10^{-7}$ Torr) under thermal evaporation. The modified tool's schematic diagram is shown on **Fig. 1**. Deposition at reduced pressure enabled us to reach the sublimation point (~130°C) before the melting point (~240°C) of DC-MBI, achieving an average deposition rate of ~0.48 A/s. These films were deposited on glass substrates with interdigitated (IDE) Pt electrodes (G-IDEPT5, Metrohm, US) with a period of 10 µm (gap of 4 µm) (**Fig. S10**). The electrodes were connected to a Radiant Precession RT66C Ferroelectric Tester system for real-time in situ electrical measurements. The deposition was carried out at room temperature and at various cryogenic temperatures controlled by a Lakeshore Temperature control system with liquid nitrogen circulation to attain cryogenic conditions. The Quartz Crystal Monitor (QCM) integrated into the deposition system continuously monitored real-time film thickness. The actual thickness was calibrated later using a couple of test deposited samples, deviating from the nominal thickness by ~100 nm. The spherulitic characteristics of the samples were assessed using a reflective cross-polarized Nikon Eclipse L200N microscope. Micro-scale topographical features were examined using a Keyence Laser Scanning Microscope, while nano-scale features were investigated using ScanAssyst Peak Force Tapping Mode on a Bruker Icon Atomic Force

Microscope (AFM). Post-processing annealing of samples in an ambient environment was conducted using a Furnace 1200 Oven. Bruker's D8 2D detector was used as the diffractometer.

**Acknowledgements:**

This research was primarily supported by the U.S. Department of Energy (DOE), Office of Science, Basic Energy Sciences (BES), under Award No. DE-SC0019173. The work is also supported in part by the Nebraska Center for Energy Sciences Research (NCESR). The research was performed in part at the Nebraska Nanoscale Facility: National Nanotechnology Coordinated Infrastructure and the Nebraska Center for Materials and Nanoscience, which are supported by the NSF under Grant No. ECCS-2025298 and the Nebraska Research Initiative.

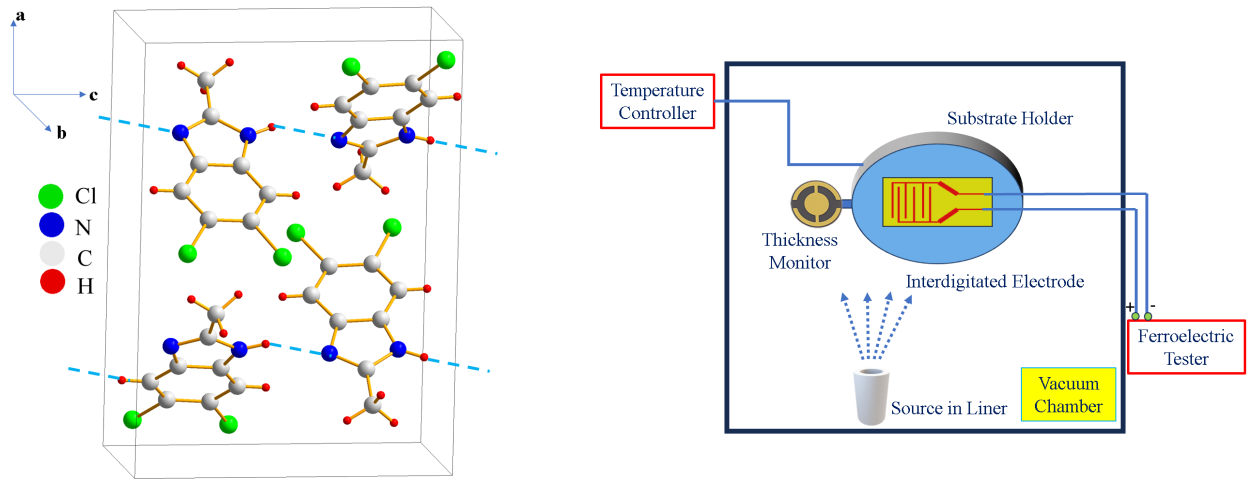

**Figure 1. Fig(a)** shows the unit cell of the DC-MBI crystal structure with hydrogen bonding pointed along c-axis shown in dotted form. **Fig(b)** shows the schematic experimental setup for the low-temperature deposition –cooling condition is maintained through the continuous supply of liquid nitrogen.

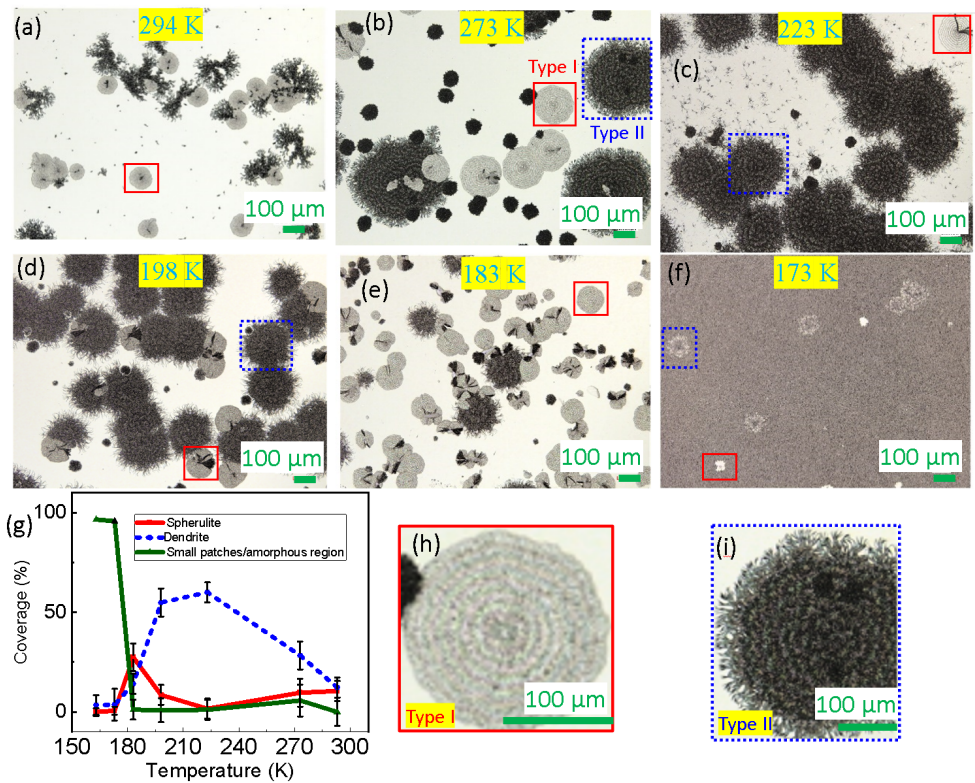

**Figure 2.** Laser Microscope images for 1μm thickness on (001) Sapphire at various substrate temperatures: 294 K (**a**), 273 K (**b**), 223 K (**c**), 198 K (**d**), 183 K (**e**), 173 K(**f**). At 163 K and below we find optical morphology like (**f**). **Fig (g)** Percentage coverage of spherulite, dendrite and other

regions for different T$_s$ samples. **Fig (h & i)** are closeup laser microscope images for Type-I and Type-II features.

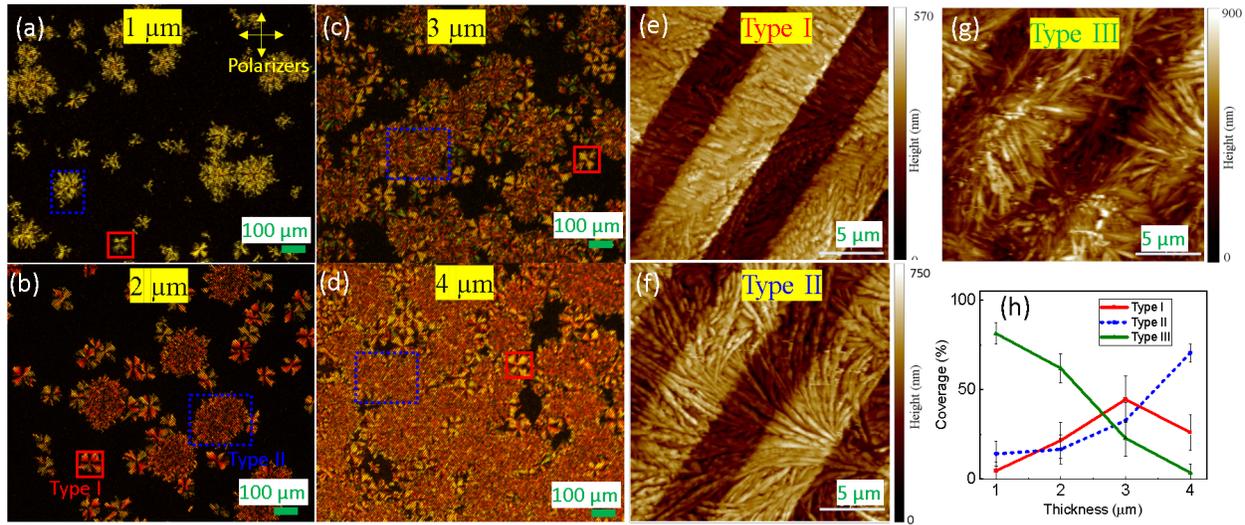

**Figure 3.** **Fig (a, b, c &d)** represents crossed polarized images over IDE substrates for thickness 1 μm, 2μm, 3μm and 4 μm respectively. **Fig (e, f &g)** shows the AFM images over the IDE for Type-I, Type-II and Type-III features. **Fig (h)** shows the thickness dependence of three typical areas labeled as I (Spherulite), II and III on **Fig 2 (b).**

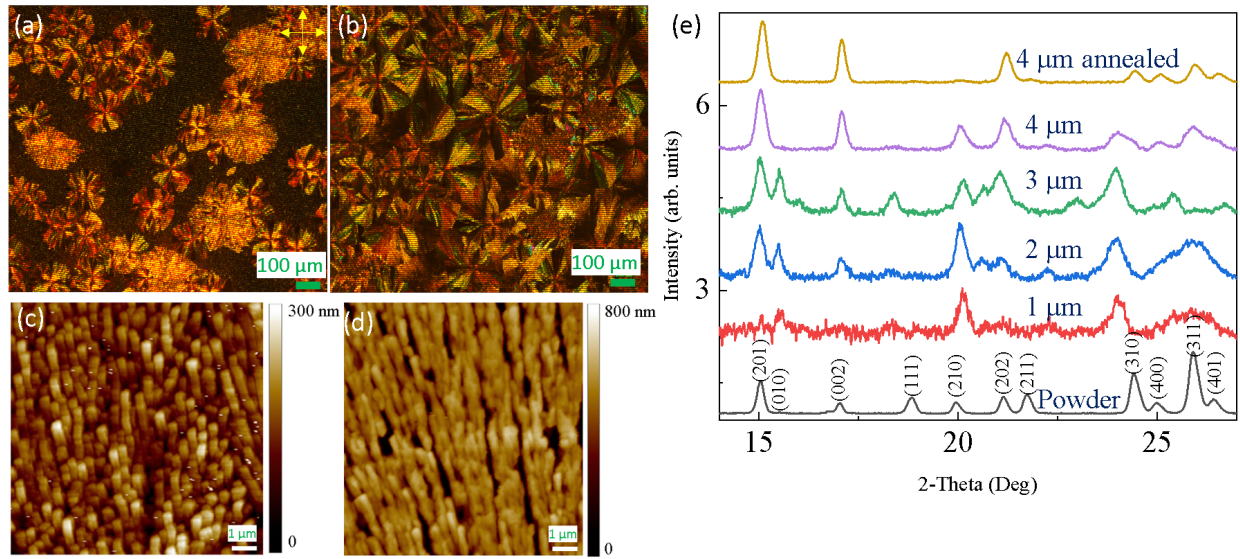

**Figure 4.** Effect of annealing DC-MBI films. **Fig (a, b)** shows cross polarized image of 4 μm grown sample on bare IDE substrate before and after annealing (not on exact same area but is typical behavior) at 80 C respectively. **Fig (c, d)** shows AFM images over the part of the spherulite before and after annealing. **(e)** XRD for the powder and as grown 1 μm, 2 μm, 3 μm, 4 μm, and 4 μm annealed samples; all data is consistently collected for 1 hr. on Bruker's D8.

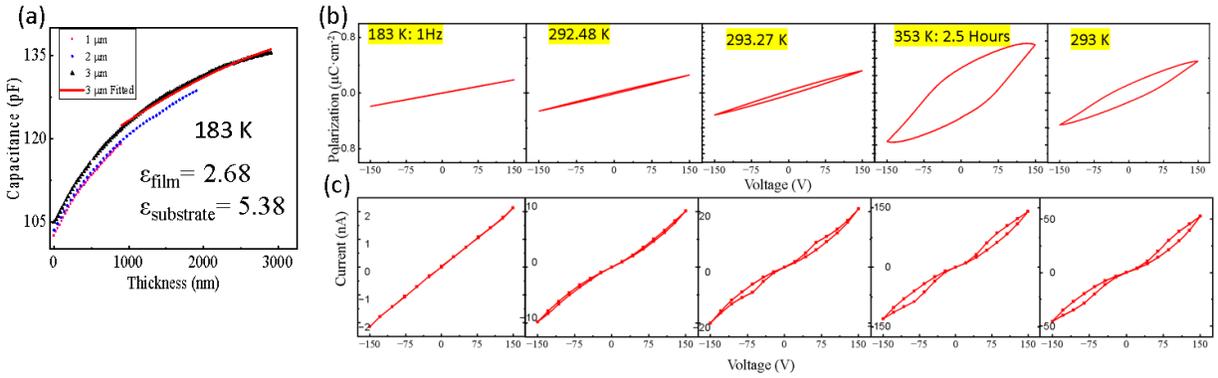

**Figure 5. (a)** Thickness dependence of IDE capacitive response at 183 K in the paraelectric phase; graph is fitted based on parallel capacitance model of IDE with the validation of necessary criterion[24]. **(b)** IDE device trend during various stages of annealing from 183 K to 353 K and back to room temperature for electrical polarization and current **(c)**.

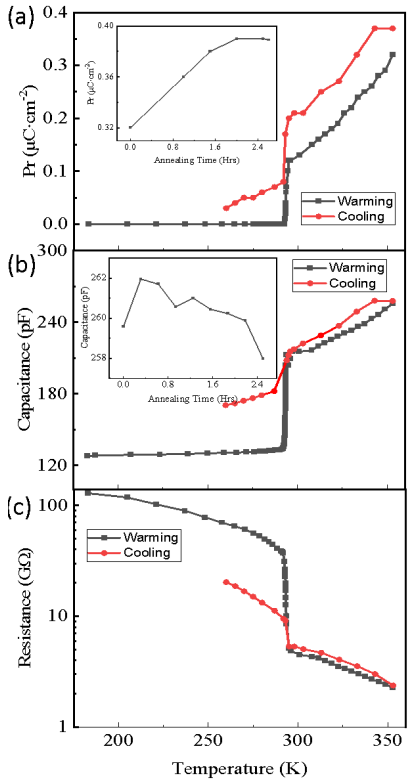

**Figure 6.** The remanent polarization (a), Capacitance (b) and Resistance (c) as a function of temperature showing the structural phase transition at 293 K. The thickness of the sample is 4 μm, and the resistance is extracted from I-V curves at 150 V.

# Spherulite-enhanced Macroscopic Polarization in Molecular Ferroelectric Films from Vacuum Deposition

## Supplementary Note 1. Characterizations for powder DC-MBI

5,6-Dichloro-2-methylbenzimidazole (DC-MBI) was brought from TCI, US. The powder sample morphologies were observed through Scanning Electron Microscope (**Fig. S1a & Fig. S1b**) and through Transmission electron microscope (**Fig. S1c**). The sample was further examined through X-ray diffraction (**Fig. S1d**) and compared to database (CCDC 909439).

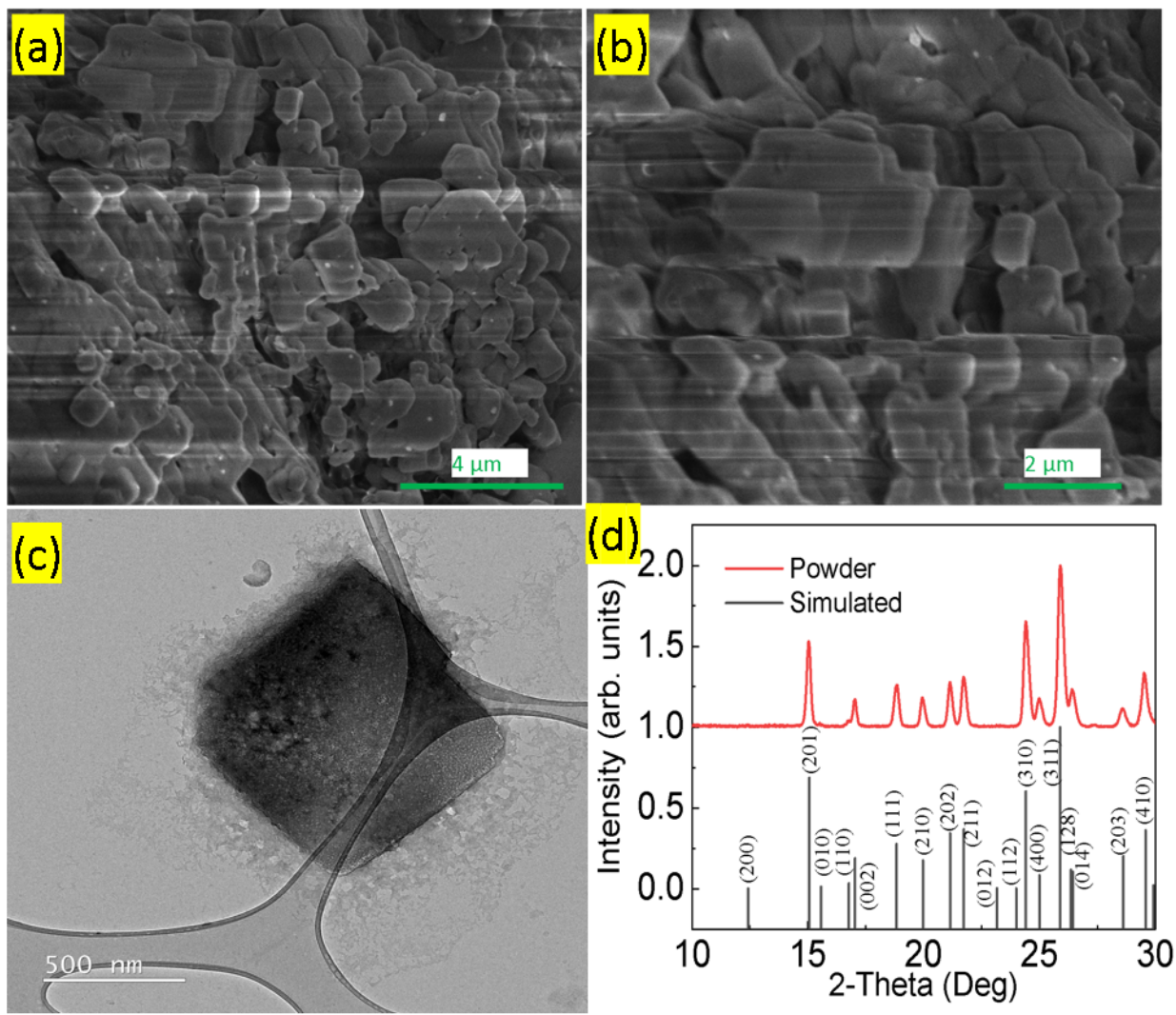

**Figure S1. Fig (a & b)** are SEM (3 kV) images at different resolutions while **Fig(c)** is TEM (200 keV) images for powder DC-MBI and **Fig(d)** is the XRD for simulated and powder sample.

**Supplementary Note 2. Characterizations for DC-MBI films**

DC-MBI films were deposited at substrate temperature $Ts$ = 183 K on various substrates. **Fig. S2** shows the closeup view of spherulite and dendrites and the corresponding Scanning Electron Microscope. SEM images clearly show the microscopic arrangements of the fibers as observed on the as grown films. The morphological distributions were similar on various substrates: **Fig. S3a and Fig. S3b** shows the cross-polarized images on gold plated Silicon substrate and IDE substrate respectively. **Fig. S4** shows qualitative comparison for the powder and films of DC-MBI as observed through X-ray Photoemission Spectroscopy (XPS). **Table 1** shows the statistics of the various types of morphologies as observed on the films grown on sapphire substrate. **Fig. S5** shows the detrimental effect on film morphologies on annealing at higher temperature suggesting for lower (353 K) annealing temperature to preserve the film morphologies and improve crystallinity.

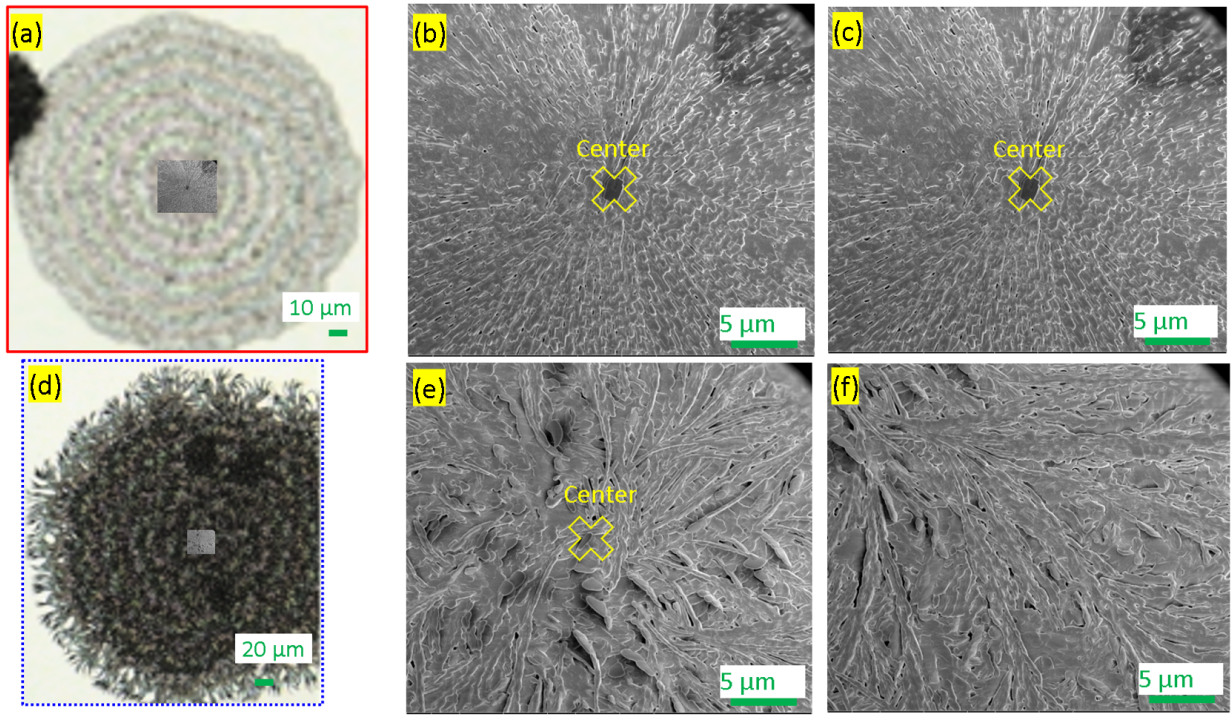

**Figure S2. Fig (a &b)** are the closeup view of Type-I and Type-II features as observed on laser microscope. **Fig (b &c)** are SEM images for Type-I while **Fig (e &f)** are corresponding SEM images for Type-II feature.

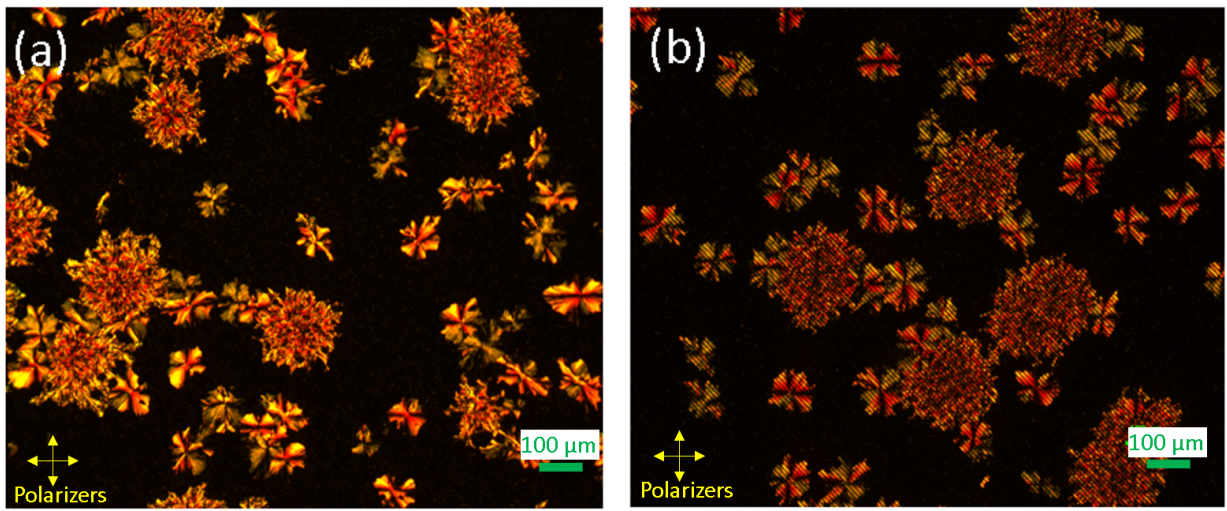

**Figure S3.** Crossed Polarized Images for the 2 μm DC-MBI films grown on gold plated silicon substrate **(a)** and Pt Interdigitated Electrodes **(b).**

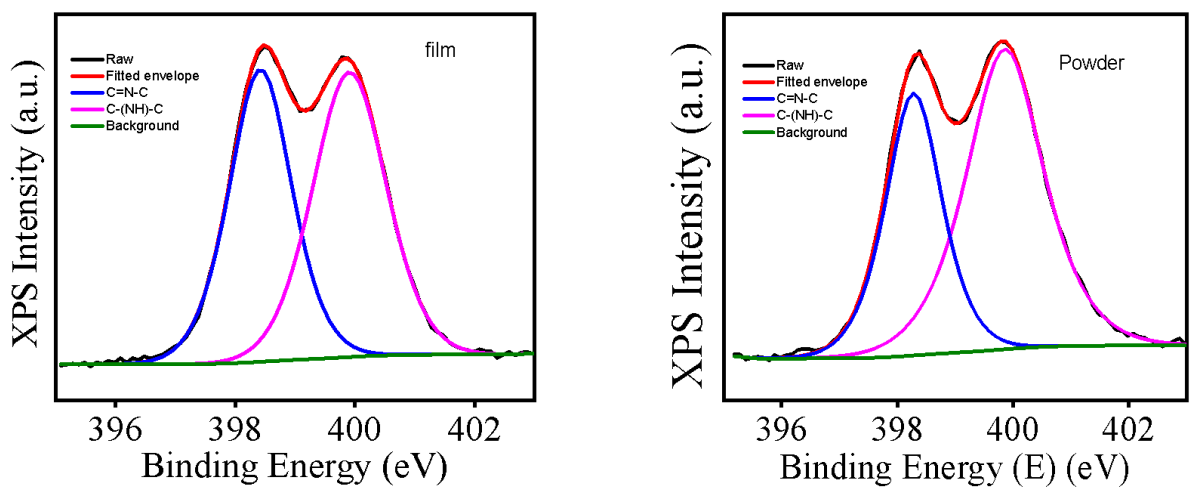

**Figure S4**. XPS data for Nitrogen Scan with peak fitting[1]. The figure on the left is for the film while that on right is for powder. Sample thickness is 4 μm.

**Table 1**. Statistics (Average height and rms roughness) of features observed on sapphire substrate for DC-MBI grown at different substrate temperatures.

| Temperature (K) | Spherulite: Type-I ($H_{avg} \pm R_q$) (μm) | Dendrite: Type-II ($H_{avg} \pm R_q$) (μm) | Black, small patches ($H_{avg} \pm R_q$) (μm) |
|---|---|---|---|
| 294 | 1 ±0.70 |  | 7 ±5.7 |
| 273 | 1±0.6 | 2±1.2 | 4±1.7 |
| 223 | 0.5± 0.5 | 2±1.3 | 8 ±4 |
| 198 | 1.5±1.0 | 8±4 | 3.8±2.1 |
| 183 | 1.4±0.85 | 4.5±3.4 | 3.7±1.2 |
| 173 | 1.4±0.6 | 2.5±1.5 | 3±2.5 |
| 163 |  | 3.5±1.5 | 3±1.0 |

## Supplementary Note 3. Electrical measurements for DC-MBI films

The DC-MBI films were deposited at $T_s$ = 183 K on IDE substrates and the corresponding electric displacement (D) and the voltage (V) graph measured for various thickness (**Fig. S5**) show the linear behavior. The local PFM measurements were conducted on the DC-MBI films deposited over the gold substrate to qualitatively demonstrate the microscopic switchable polarization. **Fig. S6** and **Fig. S7** show corresponding PFM and poling measurements respectively.

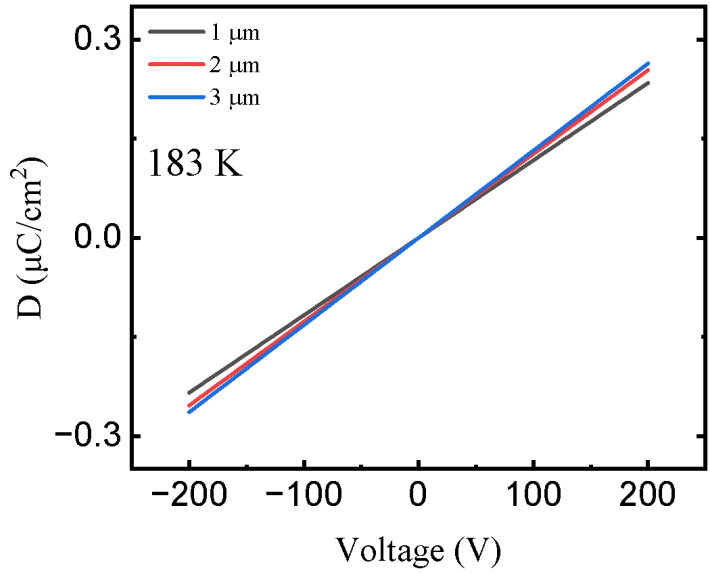

**Figure S5.** D (Electric Displacement) vs V(Voltage) for DC-MBI grown on IDE substrate at 183 K for various thickness.

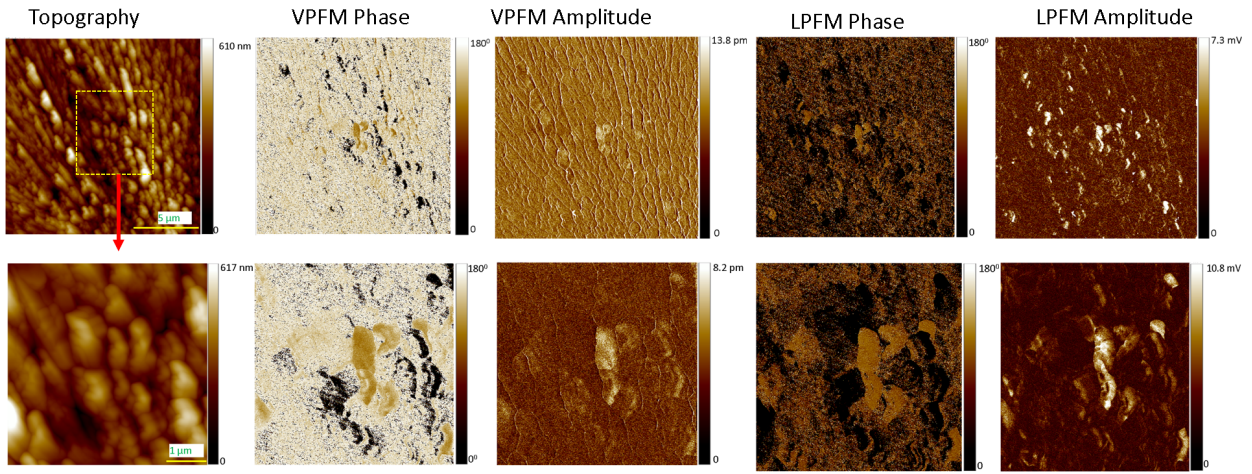

**Figure S6.** PFM measurement performed on the 1 μm thickness sample grown on Au/Si substrate at room temperature. The top row shows the corresponding Topography, VPFM Phase, VPFM Amplitude, LPFM Phase and LPFM Amplitude for 15 μm x 15 μm area while the bottom row is the zoomed over 5 μm x 5 μm. 6V ac bias is used during this scan.

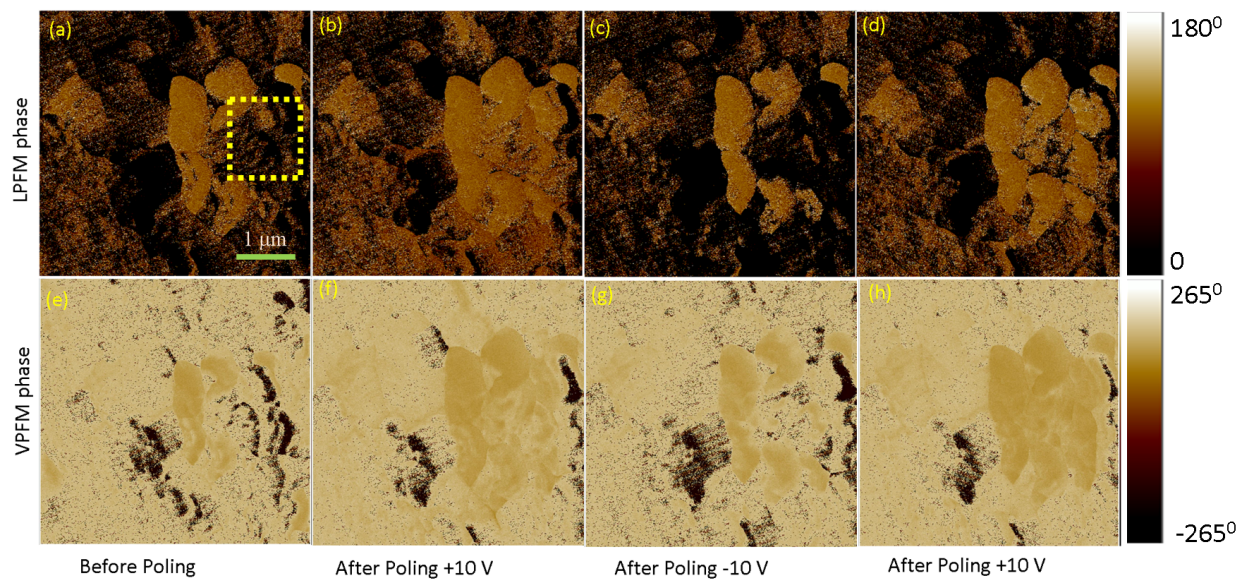

**Figure S7**. PFM Poling experiment with 10 V Tip Bias. The top row represents the LPFM over cycles of poling while the second row is the corresponding VPFM phase.

## Supplementary Note 4. Effect of annealing at higher temperature on 2D morphologies

We find that annealing at higher temperatures rather promotes the 3D growth and even dewetting and evaporation of the films causing the complete destruction of 2D morphologies.

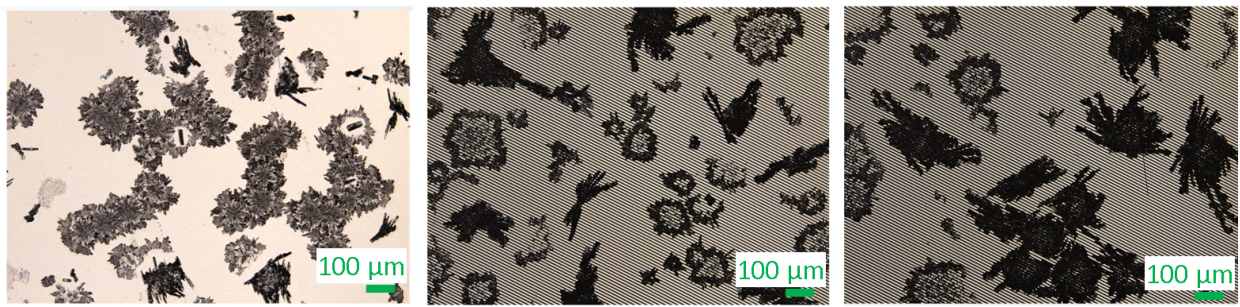

**Figure S8.** Laser microscope images of the 1 μm samples after annealing at 120 $^0$C (30 min) on sapphire substrate (**a**) over the IDE (**b**). **Fig (c)** shows the laser microscope image for the 1 μm sample annealed at 150 $^0$C (15 min) on IDE.

## Supplementary Note 5. FE/DE bilayer model

We can refer to the previously developed FE/DE bilayer model[2] to explain various subtleties of polarizations. We can relate the voids observed on the films (Type-II and Type-III features) can be qualitatively compared to dielectric. **Fig. S9b** shows how the dielectric loading on FE materials can hinder the observable polarizations.

If we use a double FE/DE double layer structure to model the porous structure, the FE loading is then[2],

$$\frac{\sigma_0}{P_0} = \frac{1}{1+\varepsilon_F\frac{1-x}{x}} = \frac{1}{1-\varepsilon_F+\frac{\varepsilon_F}{x}}, where, \ x = \frac{t_F}{t_D+t_F}.$$

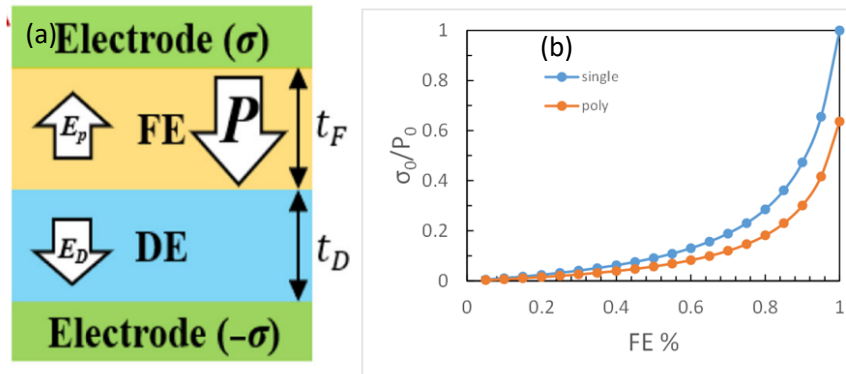

**Figure S9**. FE loading model based on FE/DE double layer structure[1](a) and measurable polarization vs ferroelectric loading the films (b).

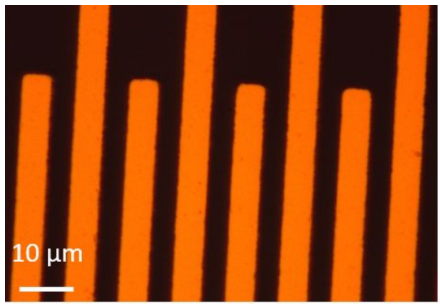

**Figure S10**. Interdigitated electrodes as viewed from optical microscope.